\documentclass[twocolumn,floatfix,prl,showpacs,superscriptaddress]{revtex4}
\usepackage{graphicx}
\begin{document}
\title{ Spin-Triplet Excitons in the $S=\frac{1}{2}$ Gapped Antiferromagnet BaCuSi$_2$O$_6$: Electron Paramagnetic Resonance Studies}
\author{S. A. Zvyagin}
\affiliation{Dresden High Magnetic Field Laboratory (HLD),
Forschungszentrum Rossendorf, Postfach 510119, D-01314 Dresden,
Germany}
\author{J. Wosnitza}
\affiliation{Dresden High Magnetic Field Laboratory (HLD),
Forschungszentrum Rossendorf, Postfach 510119, D-01314 Dresden,
Germany}
\author{J. Krzystek}
\affiliation{National High Magnetic Field Laboratory, Florida
State University, Tallahassee, FL 32310, USA}
\author{R. Stern}
\affiliation{National Institute of Chemical Physics and
Biophysics, 12618 Tallinn, Estonia}
\author{M. Jaime}
\affiliation{National High Magnetic Field Laboratory, Los Alamos
National Laboratory, MS-E536, Los Alamos, NM 87545, USA}
\author{Y. Sasago}
\affiliation{Central Research Laboratory, Hitachi Ltd.,
Higashi-Koigakubo, Kokubunji-shi, Tokyo 185-8601, Japan}
\author{K. Uchinokura}
\affiliation{RIKEN, Wako 351-0198, Japan}

\begin{abstract}
BaCuSi$_2$O$_6$, a $S=\frac{1}{2}$ quantum antiferromagnet with a
double-layer structure of Cu$^{2+}$ ions in a distorted
planar-rectangular coordination and with a dimerized spin singlet
ground state, is studied by means of the electron paramagnetic
resonance technique. It is argued that multiple absorptions
observed   at low temperatures are intimately related to a
thermally-activated spin-triplet exciton superstructure. Analysis
of the angular dependence of exciton modes in BaCuSi$_2$O$_6$
allows us to accurately estimate anisotropy parameters. In
addition, the temperature dependence of  EPR intensity and
linewidth is discussed.

\end{abstract}
 \maketitle Electron
paramagnetic resonance (EPR) is known as an extremely powerful
tool for probing the magnetic excitation spectrum in
exchange-coupled spin systems.  A new theoretical approach for
calculating low-temperature EPR parameters  of $S=\frac{1}{2}$ AFM
chains, which is based on bosonization and the standard
Feynman-Dyson self-energy formalism, has been recently developed
by Oshikawa and Affleck \cite{OshikawaAffleck-esr}. The theory
allows a precise calculation of the EPR parameters (such as
linewidth and $g$ factor shift) and their dependence on
temperature and magnetic field for uniform $S=\frac{1}{2}$ AFM
chains and chains with staggered $g$ factor and the
Dzyaloshinskii-Moriya interaction (so called sine-Gordon spin
chains). The predictions were found in excellent agreement with
experimental results \cite{Zvyagin}.

In this article we continue studying EPR properties of
low-dimensional (low-D) quantum spin systems. BaCuSi$_2$O$_6$
(barium copper cyclosilicate, also known as Han Purple Pigment
\cite{Jaime}) can be regarded as an almost ideal realization of
the $S=\frac{1}{2}$ system of weakly-interacting spin dimers with
the spin-singlet ground state and gapped excitation spectrum
\cite{Sasago}. By application of an external magnetic field the
gap can be closed, creating a gas of $interacting$ bosonic
spin-triplet excitations (triplons). In BaCuSi$_2$O$_6$ this
phenomenon can be effectively described in terms of the
field-induced Bose-Einstein condensation of triplons
\cite{Jaime,Sebastian}. Here, we focus on another interesting
phenomenon associated with interacting excited triplets  but in
the low-field quantum-disordered state, employing temperature as a
tuning parameter. We argue, that a fine structure observed in
low-temperature EPR spectra of BaCuSi$_2$O$_6$ is a fingerprint of
triplet excitations (excitons), which are mobile at low
temperatures and getting localized when temperature is increased.
We show that  anisotropic interactions in BaCuSi$_2$O$_6$ play a
critical role significantly affecting the low-temperature EPR
linewidth behavior.

\begin{figure}
%[!b]
\begin{center}
\vspace{1.5cm}
\includegraphics[width=0.45\textwidth]{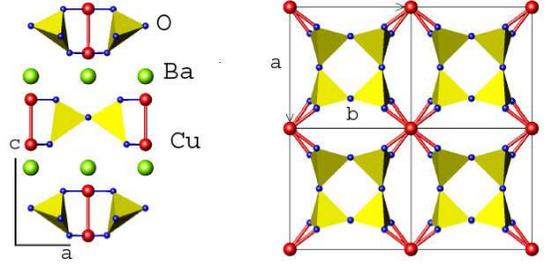}% Here is how to import EPS art
\vspace{-3.5 cm} \caption{\label{fig:BCSO-stru} Schematic views of
the room-temperature crystal structure of BaCuSi$_2$O$_6$,
composed of SiO$_4$ tetrahedra (Si atoms are located in
tetrahedron centers), CuO$_4$ distorted squares, and Ba atoms. The
Cu$^{2+}$ ions in the Cu$_2$Si$_4$O$_{12}$-layers form a
quasi-two-dimensional system of weakly-interacting dimers.}
\end{center}
\end{figure}

BaCuSi$_2$O$_6$  crystallizes in the tetragonal
non-centrosymmetric space group $I\overline{4}m2$ ($Z=4$,
$a=7.042$ ~\AA, $c=11.133$ ~\AA) \cite{Finger}. Magnetic Cu$^{2+}$
ions are arranged in layers with the $c$ axis perpendicular to the
layers (Fig.\ \ref{fig:BCSO-stru}). Within each layer  Cu$^{2+}$
ions form a distorted planar-rectangular-coordination structure of
$S=\frac{1}{2}$ sites (the O-Cu-O angle corresponding to bonds
along the rectangular diagonal direction, 177.8$^o$ \cite{Karine},
deviates from that for an ideal planar configuration).  The
Cu$^{2+}$ ions from each two neighboring layers are coupled with
each other, forming a two-dimensional dimer structure. The
intradimer distance is relatively small, 2.743~\AA, while the
distance between dimers is 7.042~\AA. The Cu bilayers are
structurally separated from each other by planes of Ba$^{2+}$
ions. The quasi-2D nature of the interdimer interactions in
BaCuSi$_2$O$_6$ has been confirmed by means of inelastic neutron
scattering \cite{Sasago}, yielding $J_1\sim51$ K  and $J_2\sim
2.2$ K  at temperature T = 3.5 K (where $J_1$ and $J_2$ are intra-
and interdimer exchange coupling constants, respectively).

\begin{figure}
%[!b]
\begin{center}
\vspace{1.5cm}
\includegraphics[width=0.45\textwidth]{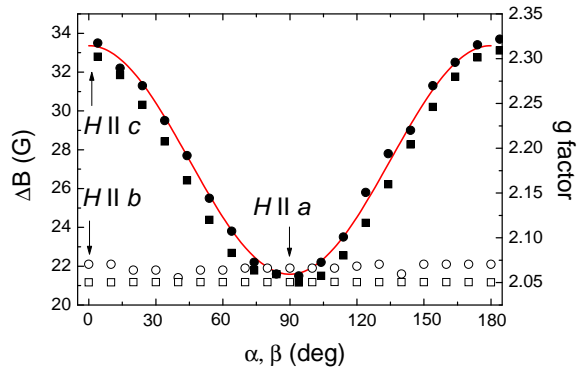}% Here is how to import EPS art
\vspace{-2.5 cm} \caption{\label{fig:BCSO-Angle} Room-temperature
angular dependence of the  $g$ factor and the peak-to-peak EPR
linewidth in BaCuSi$_2$O$_6$, $\Delta B$, in the $ac$ plane (close
squares and circles, respectively), and in the $ab$ plane (open
squares and circles, respectively). The line is a fit of the
angular dependence of the EPR linewidth in the $ac$ plane (see the
text for details).}
\end{center}
\end{figure}

A well-defined single EPR line has been observed in
BaCuSi$_2$O$_6$ at room temperature. In Fig.\
\ref{fig:BCSO-Angle},  the angular dependences of the $g$ factor
and peak-to-peak EPR linewidth, $\Delta B$, of this absorption in
the $ac$ plane, and in the $ab$ plane, are shown. The line in
Fig.\ \ref{fig:BCSO-Angle} corresponds to a fit of the angular
dependence of the EPR linewidth in the $ac$ plane using the
standard formula
\begin{eqnarray}
\label{Angle} \Delta B/\Delta B _{\perp}=1+c~\mbox{cos}^2 \alpha,
\end{eqnarray}
where $\Delta B$ is the peak-to-peak linewidth and $c \sim$ 0.54.
Both the linewidth and $g$ factor are anisotropic within the $ac$
plane with a maximum along the $c$ axis ($\Delta B_{\parallel}$ =
33.5 $\pm$ 0.5 G, $g_{\parallel}$ = 2.306 $\pm~0.003$), while
being minimal and isotropic in the $ab$ plane ($\Delta B_{\perp}$
= 21.7 $\pm$ 0.5 G, $g_{\perp}$ = 2.050 $\pm~0.003$). Since
$g_{\parallel}\neq$ 2 and $g_{\parallel}
> g_{\perp}$  the orbital singlet state $d_{x^2-y^2}$ can be
identified  as the ground state, with electron-density maxima
located in the basal $ab$ plane \cite{Abragam}.
 Such a configuration suggests the absence of direct
orbital overlaps between Cu sites from  neighboring layers,
revealing a strong superexchange mediating the Cu-Cu intradimer
coupling.

\begin{figure}
%[!b]
\begin{center}
\vspace{2.5cm}
\includegraphics[width=0.45\textwidth]{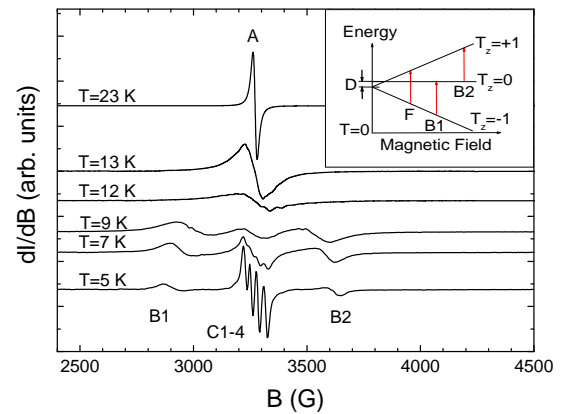}% Here is how to import EPS art
\vspace{-2.5 cm} \caption{\label{fig:BCSO-LT-Spectrum} The
evolution of the low-temperature EPR spectra of BaCuSi$_2$O$_6$ at
a frequency of 9.385 GHz; the magnetic field is applied along the
$a$ axis.  Note that points at lower temperatures are taken with
an increased amplification rate.  The lines A, B1 and B2
correspond to  spin-triplet excitons. The hyperfine structure C1-4
is related to  interactions within unpaired doublet Cu$^{2+}$
sites, caused by defects. The line F (Fig.\
\ref{fig:BCSO-LT-Angle}) is outside the shown field range. The
inset shows  a schematic view of the energy-level diagram for
dimerized $S=\frac{1}{2}$ systems in the tetragonal
configuration.}
\end{center}
\end{figure}

\begin{figure}
%[!b]
\begin{center}
\vspace{2 cm}
\includegraphics[width=0.45\textwidth]{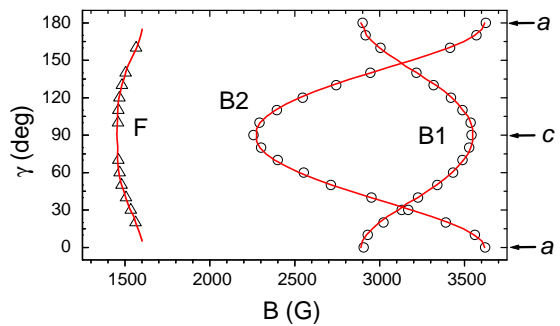}% Here is how to import EPS art
\vspace{-3 cm} \caption{\label{fig:BCSO-LT-Angle} Angular
dependence of the EPR spectrum in BaCuSi$_2$O$_6$ in the $ac$
plane, taken at a frequency of 9.385 GHz and a temperature of 6 K.
$\gamma$ is the angle between the direction of the applied field
and the $a$ axis. Triangles and circles correspond to experimental
data, while lines correspond to stimulation results (see text for
details).}
\end{center}
\end{figure}

In contrast to the room-temperature data, a very rich EPR spectrum
has been observed in BaCuSi$_2$O$_6$ at low temperatures (Fig.\
\ref{fig:BCSO-LT-Spectrum}). The central line is split into four
components labelled C1-4, which are clearly a sign of the
hyperfine structure due to the I = 3/2 nuclear spin of unpaired
Cu$^{2+}$ sites,  caused by defects. A difference between
resonance fields  of absorptions B2 and B1 corresponds to a
splitting between T$_z$ = 0 and T$_z$ = $\pm$ 1 excited states
(Fig.\ \ref{fig:BCSO-LT-Spectrum}, inset).
 The absorption  F (not shown in Fig.\
\ref{fig:BCSO-LT-Spectrum}, but plotted in Fig.\
\ref{fig:BCSO-LT-Angle}) is the so-called ``half-field'' resonance
and corresponds to transitions between T$_z$ = -1 and T$_z$ = +1
levels. Since $\Delta M_{S}$ = 2, these transitions are nominally
forbidden, but become allowed by symmetry in certain situations,
e.g.,  when wave functions from neighboring spin levels are mixed
\cite{Abragam}.

Studying the angular dependence of the lines B1 and B2 allows an
approximate estimate of the anisotropy parameter $D$ using the
expression
\begin{eqnarray}
\label{R-Field} B=B_0 \pm \frac{1}{2} \frac{D}{g \mu_B} (3
\frac{g_{\parallel}^2}{g^2}\mbox{cos}^2 \Theta -1),
\end{eqnarray}
where $g^2=g_{\parallel}^2 \mbox{cos}^2 \Theta + g_{\perp} ^2
\mbox{sin}^2 \Theta$ \cite{Abragam}. In our calculations we used
the approach developed by Baranowski $et~al.$ \cite{Baranowski},
which allows for a more accurate estimate of the zero-field
splitting parameters for a fully anisotropic $g$ tensor and for
every orientation of the field $B$. Results of the simulation are
shown in Fig.\ \ref{fig:BCSO-LT-Angle} by lines. The best fit
yields $g_{\perp}=2.05$, $g_{\parallel}=2.31$ for the $g$ factor,
and $D$ = 740 G (0.1 K). Interestingly, the $g$ factors obtained
from low-temperature data simulations are consistent with those
obtained from room-temperature experiments.

\begin{figure}
%[!b]
\begin{center}
\vspace{2 cm}
\includegraphics[width=0.45\textwidth]{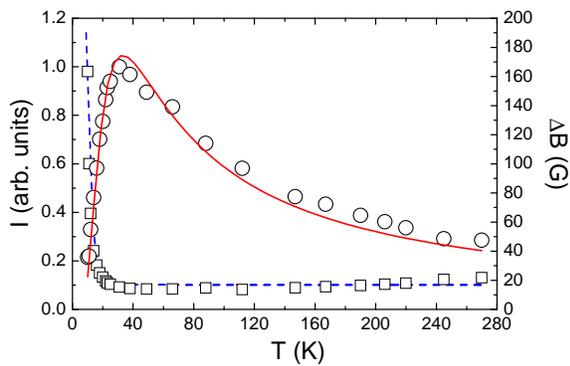}% Here is how to import EPS art
\vspace{-2.5 cm} \caption{\label{fig:BCSO-TD} Temperature
dependencies of the integrated EPR intensity $I$ (circles) and
peak-to-peak linewidth
 $\Delta B$ (squares), measured with magnetic field $H$ parallel to the $a$
axis at temperatures down to 10 K. The solid line is a fit using
an isolated dimer model with an energy gap $\Delta$ = 53 K. The
dashed line is a guide for eye. }
\end{center}
\end{figure}

Upon warming up the transitions B1 and B2 move towards each other
and merge (Fig.\ \ref{fig:BCSO-LT-Spectrum}), so that only one
absorption line A can be resolved at temperatures higher than
$\sim$ 12 K. Spin-triplet excitations contribute to the EPR
intensity, resulting in a pronounced peak in the temperature
dependence at $\sim$ 40 K (Fig.\ \ref{fig:BCSO-TD}, circles).
Since the interdimer exchange interaction in BaCuSi$_2$O$_6$ is
significantly weaker than the intradimer coupling ($J_2\ll J_1$),
the isolated dimer model is used to fit the EPR data. The
Boltzmann distribution is included to analyze the temperature
dependence of the EPR signal and to determine the singlet-triplet
energy gap $\Delta$. Normally,  the EPR intensity is determined by
a difference between populations of the lowest excited states, and
at sufficiently low temperatures ($T\ll\Delta$) should be
proportional to the number of thermally-activated triplet states.
Thus, for the integrated absorption (and assuming a temperature
independent size of the energy gap) we can write $I(T)\sim\{\exp
[(-\Delta +g\mu_B B)/k_BT] - \exp [(-\Delta -g\mu_B B)/k_BT]\}/Z$,
 where  $\mu_B$ is the Bohr magneton, $k_B$
is the Boltzmann constant, $Z$ is the partition function for the
singlet-triplet energy scheme, and $\Delta$ is the singlet-triplet
gap. The best fit of the experimental data yields the mean energy
gap
 of $\Delta$ = 53 K, which is in excellent agreement with the data
obtained from the neutron scattering  \cite{Sasago}.

 Since their relevancy to the understanding EPR properties of BaCuSi$_2$O$_6$,
let us now discuss results of  neutron scattering measurements
\cite{Sasago} in  more detail. It was observed that at low
temperatures the interdimer coupling results in a relatively small
but yet pronounced dispersion of triplet excitations along the
layer direction. The physical picture of the dispersion of
magnetic excitations in BaCuSi$_2$O$_6$ can be associated with
single-dimer (local) excitations ``hopping'' from one site to
another within the Cu$^{2+}$ bilayers.  The delocalized,
``excitonic'' nature of spin-triplet excitations has been
revealed, for instance, in some organic charge-transfer salts with
spin singlet ground state \cite{Krzystek}. It was shown that the
exciton motion can be particularly prominent in low-D materials:
it is very fast along one crystallographic direction and slow
along other directions (1D propagation), or very fast within one
crystallographic plane (2D propagation).   The growing number of
excitons causes pronounced changes in the dispersion of magnetic
excitations in BaCuSi$_2$O$_6$. As follows from neutron-scattering
measurements \cite{Sasago}, at T $\sim$ 50 K the bandwidth along
the in-plane direction is significantly reduced compared to that
at 3.5 K, so that the dispersion of the triplet excitation becomes
almost flat at high temperatures. Qualitatively the observed
behavior can be described as following: at high temperatures
excitons can not freely hop form one site to another (since most
of the sites are already occupied), so that the dimer excitations
have to reside on their own sites. Thus,  triplets get localized,
which manifests itself in the suppression of the excitation
dispersion at high temperatures.

Let us now discuss the EPR results. As we saw above, the fine
structure observed in the low-temperature regime can be nicely
reproduced taking into account an anisotropy parameter $D$ = 0.1
K, characterizing the excited triplet state. Upon warming up, the
two peaks in the fine structure are getting closer to each other.
At $T\sim$ 12 K (depending on the orientation, i.e., the original
splitting of the two lines) the fine structure can not be resolved
any more, and a single line is observed. Upon further temperature
increase, the EPR line is getting  narrower (Fig.\
\ref{fig:BCSO-TD}) until the linewidth becomes almost constant at
temperatures higher than $T\sim$ 20 K. All these phenomena are
textbook examples of the temperature-activated exchange processes
taking place between the particular dimers excited to the triplet
state. As described e.g. in Ref. \cite{Carrington},   in the
slow-exchange regime (exchange frequency $\nu \ll g\mu_B|B2 -
B1|$) the two lines corresponding to the individual $\Delta M_{S}$
= 1 transitions get broadened upon raising the temperature, and
start moving towards each other, while in the fast exchange regime
($g\mu_B|B2 - B1| \ll \nu $) the single averaged line gets
narrowed with increasing temperature \cite{note}.

The nature of the exchange cannot be easily discerned from the
EPR-detected effects alone. There are two possibilities, which can
be termed ``static exchange'' and ``motional exchange''. In the
first case the triplet excitations are immobile, while in the
second case they move in-plane, ``bumping''  into each other with
a frequency dependent on the concentration of the excited dimers.
Combining the EPR observations with those obtained from neutron
scattering one can clearly decide that it is the ``motional
exchange'' that takes place in the system up to about 20 K. Above
that temperature most of the dimers are already excited into the
triplet state, and the excitons become immobilized. In such a
case, the ``static exchange'' takes over, which is clearly not
thermally activated.

The angular dependence of the EPR linewidth in the fast exchange
regime requires some consideration. If the anisotropy of the EPR
spectrum as shown in  Fig.\ \ref{fig:BCSO-Angle} were related
entirely to the underlying fine structure, then the linewidth
minima should appear at the two crossover points in the
slow-exchange spectra - as shown in Fig.\ \ref{fig:BCSO-LT-Angle},
i.e., approximately 30$^o$ from the $a$ axis. Since this is not
the case, with the linewidth minimum appearing along the $a$ axis,
and the linewidth following Eq.\ \ref{Angle}, it means that the
reasons for the linewidth anisotropy must be more complex.
Generally speaking, the EPR linewidth in low-D spin systems can be
affected by several factors, including symmetric term (anisotropic
exchange and  dipole-dipole interactions) and the
Dzyaloshinskii-Moriya interaction (for most recent discussion, see
for instance,  Ref. \cite{Choukroun}). Analysis of all these
parameters in connection with the temperature dependence of the
EPR line in BaCuSi$_2$O$_6$ can be a rather complex and
challenging problem. It would be very interesting to study
theoretically a contribution  of excitons to the EPR linewidth,
whose concentration and mobility are temperature dependent.

As shown, analyzing the angular dependence of exciton modes in
BaCuSi$_2$O$_6$ allowed  the accurate determination of the
anisotropy parameter $D$. The described procedure might be of
particular importance for studying a generic phase diagram
predicted for anisotropic $S=\frac{1}{2}$ quantum chains (see, for
instance, Ref. \cite{Somma}), and can be applied for a large
number of $S=\frac{1}{2}$ quantum antiferromagnets with the
dimerized spin singlet ground state.  A low-temperature fine
structure similar to that in BaCuSi$_2$O$_6$, has been recently
revealed in the $S=\frac{1}{2}$ dimer spin system
VOHPO$_4\cdot\frac{1}{2}$H$_2$O \cite{Cao}. A multiple-peak EPR
fine structure, caused by a rather complex topology of magnetic
interactions (there are at least two non-equal positions of
Cu$^{2+}$ ions) has been observed \cite{Zvyagin-Seb} in the
quasi-2D gapped spin system $\beta$-Sr$_2$Cu(BO$_3$)$_2$
\cite{Sebastian2}. In both cases the linewidth exhibited a
temperature dependence, similar to that in BaCuSi$_2$O$_6$.
Anisotropy effects revealed in BaCuSi$_2$O$_6$, can be also
relevant to the low-temperature EPR line broadening observed in
the low-D spin antiferromagnets CuGeO$_3$ \cite{Eremina} and
$\alpha '$-NaV$_2$O$_5$ \cite{Lohmann}.

In conclusion, resonance properties of BaCuSi$_2$O$_6$, a quasi-2D
gapped antiferromagnet with  dimerized spin singlet  ground state,
has been studied by means of X-band EPR technique. We argued that
multiple absorptions observed in  low-temperature EPR spectra were
intimately related to a thermally-activated spin-triplet exciton
superstructure. We demonstrated that   the analysis of the angular
dependence of exciton modes could be used as a powerful tool for
accurate estimation of zero-field splittings within
 triplet states (and corresponding anisotropy parameters) in gapped $S=\frac{1}{2}$ quantum
antiferromagnets with the dimerized spin singlet ground state. We
hope that our experimental findings will stimulate a theoretical
interest, which would make not only qualitative, but also
quantitative interpretation of the temperature dependence of EPR
parameters in BaCuSi$_2$O$_6$ and  related materials possible.

The authors express their sincere thanks to  A.K. Kolezhuk, V.
Kataev and G. Teitel'baum for fruitful discussions. A. Ozarowski
is acknowledged for providing a spectral simulation program. The
work was supported by the Saxon Ministry of Science and the Fine
Arts (SMWK) and the Federal Ministry of Education and Research
(BMBF) of Germany. A portion of this work was performed at the
National High Magnetic Field Laboratory, which is supported by NSF
Cooperative Agreement No. DMR-0084173, by the State of Florida,
and by the DOE. S.A.Z. and R.S. acknowledge the support from the
NHMFL through the Visiting Scientist Program No. 1368 and 1314,
respectively. R.S. was supported by the Estonian Science
Foundation.

\end{document}